\documentclass[9pt]{article}
\usepackage{jhep-mod}
\usepackage{amssymb,amsmath,amsthm}
\usepackage{mathrsfs}
\usepackage{mathtools}
\usepackage[applemac]{inputenc}
\usepackage[thinc]{esdiff}
\usepackage{enumerate}
\usepackage{enumitem}

\usepackage{changepage}

\usepackage{ulem}

\usepackage[dvipsnames]{xcolor}
\usepackage{graphicx}
\usepackage{dcolumn}
\usepackage{bm}
\usepackage{hyperref}

\hypersetup{
	colorlinks=true,
	linkcolor=black!30!blue,
	citecolor=black!30!blue,
	filecolor=black,
	urlcolor=black!45!blue,
}

\newcommand{\orcidicon}{\includegraphics[width=0.32cm]{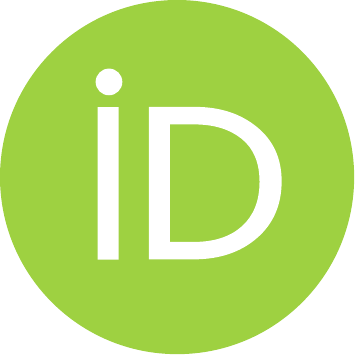}}

\newcommand\orcidT{{\href{https://orcid.org/0000-0001-8007-5181}{\orcidicon}}}
\newcommand\orcidM{{\href{https://orcid.org/0000-0002-1529-1889}{\orcidicon}}}
\newcommand\orcidP{{\href{https://orcid.org/0000-0001-8073-4896}{\orcidicon}}}

\begin{document}
	
	\title{Phantom attractors in Kinetic Gravity Braiding theories: a dynamical system approach}
	
	\author{Teodor Borislavov Vasilev\,$^{a}$\orcidT, Mariam Bouhmadi-López\,$^{b,c}$\orcidM \ {\rm and} Prado Martín-Moruno\,$^a$\orcidP}
	
	\affiliation{$^a$ Departamento de F\'isica Te\'orica and IPARCOS, Universidad Complutense de Madrid, E-28040 Madrid, Spain.}

	\affiliation{$^b$
		IKERBASQUE, Basque Foundation for Science, 48011, Bilbao, Spain
	}%
	\affiliation{$^c$
		Department of Physics and EHU Quantum Center, University of the Basque Country, UPV/EHU, P.O. Box 644, 48080 Bilbao, Spain.
	}

	\emailAdd{teodorbo@ucm.es}
	\emailAdd{mariam.bouhmadi@ehu.eus}
	\emailAdd{pradomm@ucm.es}
	
	
	\abstract{ We revise the expansion history of the scalar field theories known as Kinetic Gravity Braiding. These theories are well-known for the possibility of driving the expansion of the cosmos towards a future self-tuning de Sitter state when the corresponding Lagrangian is invariant under constant shifts in the scalar field. Nevertheless, this is not the only possible future fate of these shift-symmetric models. Using a dynamical system formulation we show that future cosmological singularities can also appear in this framework. Moreover, we present explicit examples where the future attractor in the configuration space of the theory corresponds to a big rip singularity.
		
	}
	
	\maketitle
	
	
	\section{Introduction}
	
	Scalar field theories known as Kinetic Gravity Braiding (KGB) \cite{KGB1} may provide a prominent underlying framework for describing dark energy (DE) without invoking a cosmological constant.  Indeed, these models have already proven to be extremely fruitful in both early- and late-time cosmology; see, for instance, applications to inflation in references \cite{Armendariz-Picon:1999hyi,Burrage:2010cu,Mizuno:2010ag,Creminelli:2010ba,Kobayashi:2011nu} and DE models \cite{KGB1,Tsujikawa:2010zza,Bernardo:2021hrz,Germani:2017pwt,Martin-Moruno:2015bda,Martin-Moruno:2015kaa,Abramo:2005be,Tsyba:2010ji,Linder:2021est,Panpanich:2021lsd,DeFelice:2010pv,DeFelice:2011bh}. The KGB models are a subclass of the more general Horndeski theory \cite{Horndeski:1974wa} (se also reference \cite{Kobayashi:2019hrl} for a review), and, therefore, they render second order field equations. 
	Furthermore, the KGB model trivially allows gravitational waves to propagate at the speed of light \cite{Horndeski:1974wa}, which is in agreement with the recent observation of the GW170817 event \cite{LIGOScientific:2017zic}. In addition, the parameter-space of the theory has also been confronted with cosmological observables \cite{SpurioMancini:2019rxy,Kreisch:2017uet,Noller:2018wyv,Baker:2020apq,Traykova:2021hbr}, rendering this set-up as a viable DE model. Nevertheless, note that some specific subclasses of the KGB theory may be found at tension with cosmological data (see, for example, the discussion on Cubic Galilean gravity in references \cite{Barreira:2014jha,Renk:2017rzu,Peirone:2017vcq}). The KGB set-up has also been studied in the context of the $H_0$ tension, showing a possible modest increase in the value of $H_0$ in these theories \cite{Lee:2022cyh}. Moreover, the Palatini version of the KGB theory and its connection to the metric formalism have been previously explored in reference \cite{Helpin:2019kcq}.
	
	A remarkably interesting application of the KGB framework for modelling DE is that provided by the shift-symmetric sector of the theory. That is when the KGB's action is invariant under constant shifts in the scalar field, i.e $\phi\to\phi+c$ being $c$ a constant. These shift-symmetric KGB models are well-known for the possibility of driving the expansion of the cosmos towards a future self-tuning de Sitter (dS) state \cite{KGB1}. Consequently, they have naturally attracted considerable attention  (see, for instance, references \cite{DeFelice:2010pv,Bernardo:2021hrz,Germani:2017pwt,Martin-Moruno:2015bda,Martin-Moruno:2015kaa,DeFelice:2011bh,Tsujikawa:2010zza}). In addition, the effective DE component obtained in this fashion can exhibit phantom behaviour that is stable at first order in perturbation theory \cite{KGB1}, i.e. free from ghost and gradient instabilities.  (Recall that phantom DE is characterized by an equation of state parameter, that is the ratio between the pressure and the energy density of DE, $w_{\textup{DE}}$ less than -1.) Please note that phantom DE was analytically shown to be a prerequisite for alleviating both the $H_0$ and $\sigma_8$ tensions simultaneously \cite{Heisenberg:2022gqk,Heisenberg:2022lob}. (For a discussion on the $H_0$ tension in the KGB set-up see, for instance, reference \cite{Lee:2022cyh} and references therein.)

	Nevertheless, it is a general property of phantom DE that the evolution of the universe could entail a future cosmological singularity.
	All bounded structure and, ultimately, space-time itself could be ripped apart at a final big rip (BR) singularity \cite{Starobinsky:1999yw,BR}.
	This fatal event is characterized by the divergence at a finite cosmic time of the size of the observable universe, the Hubble rate and its cosmic time derivative. A phantom dominated universe could also reach a big freeze (BF) singularity \cite{BF1BouhmadiLopez:2006fu,BF2BouhmadiLopez:2007qb}. Like in the BR scenario, the Hubble rate and its cosmic time derivative diverge in finite cosmic time but for a finite value of the scale factor. Another example of finite-size singularity is that of sudden singularity \cite{sudden1}. At this event, the Hubble rate remains finite but its cosmic time derivative explode (or some higher order derivative for the case of generalized sudden singularity \cite{sudden2}). Moreover, this finite-scale-factor singularity could take place at a finite cosmic time.
	(See references \cite{Dabrowski:2014fha,NojiriClassification,SingularityClasification,Fernandez-Jambrina:2021foi} for other examples of cosmological singularities.) Therefore, since the future phenomenology of (phantom) DE models could encompass a broad variety of possibilities, it is natural to wonder whether the (phantom) DE component modelled by the shift-symmetric KGB theories could lead the evolution of the cosmos towards a different future state from that of the well-studied dS future solution of the theory. In other words, the question arises whether a future dS is the only possible attractor in the configuration space of the shift-symmetric KGB theory.
	
	In this work we address the latter question. By reviewing the assumptions underlying the existence of these future dS attractors we argue for the possibility of different future evolutions for the system. 
	In order to support these claims, we propose a dynamical system formulation for the KGB theory different from the previously used in the literature (see, for example, references \cite{DeFelice:2010pv,Germani:2017pwt,Martin-Moruno:2015kaa,DeFelice:2011bh,Tsujikawa:2010zza}). Within this new approach, we study the fixed points of the system and their stability. 
	Please note that the existence of future singularities in the shift-symmetric KGB models was also addressed in reference \cite{TMPletter}. We presented there a simple KGB model featuring a future BR singularity and discussed its expansion history. In the present work we explore in more depth and generality the dynamical system representation of these shift-symmetric theories and we apply our analysis to different KGB models.
	
	This work is organized as follows: Section \ref{sec:KGB} provides an introduction to the shift-symmetric KGB theories and their application to a homogenous and isotropic cosmological background. Section \ref{sec:Dynamical Systems} is devoted to the dynamical system formulation of an expanding universe in the shift-symmetric KGB set-up. Moreover, different power-law KGB models are analysed in sections \ref{sec:powerLaws} and \ref{sec:Mod3}. Lastly, concluding remarks shall be found in section \ref{sec:conclusions}. Appendices \ref{app:functions} and \ref{app:BR} contain clarification notes.

	
	\section{Kinetic gravity braiding\label{sec:KGB}}
	
	The KGB theory is given by the action \cite{KGB1}
	\begin{eqnarray}\label{eq:actionKGB}
		S=\int d^4 x\sqrt{-g}\left[\frac12 R+ K(\phi,X)-G(\phi,X)\Box\phi \right],
	\end{eqnarray}
	where we have adopted the geometric unit system $8\pi G=c=1$, $K(\phi,X)$ and $G(\phi,X)$ are arbitrary functions of the scalar field $\phi$ and its canonical kinetic term $X\coloneqq-\frac12 g^{\mu\nu}\nabla_\mu\phi\nabla_\nu\phi$, and
	the box represents the covariant d'Alembertian operator $\Box \phi=g^{\mu\nu}\nabla_\mu\nabla_\nu\phi$.  The presence of this operator in action (\ref{eq:actionKGB}) introduces a mixing between the kinetic term of the metric and that of the scalar field (symbolically $G\partial g\partial \phi$). This \textit{kinetic braiding} leads to the presence of second order derivatives of the metric in the scalar field equation of motion, and vice-versa \cite{KGB1}. Moreover, the d'Alembertian operator may give rise to deviations from the perfect fluid description for the scalar field \cite{KGB1} (see also reference \cite{Pujolas:2011he}). Nevertheless, the perfect fluid form can be safely assumed for a homogeneous and isotropic cosmological background 
	\cite{Nucamendi:2019uen}.
  	
  	A remarkably interesting application of KGB models to cosmology is that provided by the shift-symmetric sector of this theory \cite{KGB1}. That is when the action (\ref{eq:actionKGB}) is invariant under the shift
  	\begin{eqnarray}
  			\phi\to\phi+c,
  	\end{eqnarray} 
	being $c$ a constant. In practise, this	implies that the functions $K$ and $G$ do not depend on $\phi$. In that case, the scalar field equation is given by the conservation of the corresponding shift-current \cite{KGB1}. 

  	
  	From now on, we will restrict our analysis to the spatially flat cosmological background described by the homogeneous and isotropic Friedmann-Lema\^{i}tre-Robertson-Walker (FLRW) line element
  	\begin{eqnarray}
  		ds^2=-N^2(t)dt^2+a^2(t)dx_3^2,
  	\end{eqnarray}
  	where $N$ is the lapse function, $a$ stands for the scale factor and $dx_3^2$ are the spatial three-dimensional Euclidean sections. The Friedmann and Raychaudhuri equations, then, read \cite{KGB1}
    \begin{align}
  		3H^2=&\rho_m+\rho_r-K+\dot{\phi}J,\label{FE1}\\
  		\dot{H}=&-\frac12\left(\rho_m+\frac43\rho_r\right)+XG_X \ddot{\phi}-\frac12\dot{\phi}J,\label{FE2}
    \end{align}
  	respectively, where $J$ is the only non-vanishing component of the shift-current. That component is \cite{KGB1}
  	\begin{align}\label{eq:J}
  		J\coloneqq \dot{\phi}K_X+6HXG_X.
  	\end{align}
  	In addition, matter and radiation have been considered as external sources to action (\ref{eq:actionKGB}). Their field equations read
  	\begin{align}
  		\dot{\rho}_m&=-3H\rho_m,\label{eq:conservationDust}\\
  		\dot{\rho}_r&=-4H\rho_r\label{eq:conservationRad},
  	\end{align} 
  	being $\rho_m$ and $\rho_r$ the energy densities for matter and radiation, respectively.
  	The evolution equation for the scalar field is given by the conservation of the  shift-current (\ref{eq:J}). On a FLRW background, this is \cite{KGB1}
  	\begin{eqnarray}
  		\frac{1}{a^3}\diff{\left(a^3J\right)}{t}=0.\label{eq:phi}
    \end{eqnarray}	
  	Hence, it is straightforward to find a first integral of motion for the scalar field \cite{KGB1}. That is
  	\begin{eqnarray}\label{eq:Jsym}
  		J=Q_0\left(\frac{a}{a_0}\right)^{-3},
  	\end{eqnarray}
  	being $Q_0$ the scalar charge associated with the shift symmetry and $a_0$ the current value of the scale factor. 
  	Equation (\ref{eq:Jsym}) implies that $J$ is either trivial, that is if and only if $Q_0=0$, or vanishes asymptotically for infinitely expanding universe. As a result,  the vanishing of this shift-current can be used to extract information about the future evolution of the theory \cite{KGB1}. It should be mentioned, however, that $J=0$ does not represent a proper fixed point of the system but a surface in the corresponding configuration space. This is because equations (\ref{FE1}), (\ref{FE2}), (\ref{eq:conservationDust}), (\ref{eq:conservationRad}) and  (\ref{eq:phi}) render a three-dimensional phase-space as they represent four dynamical equations and one constraint. Consequently, $J=0$ defines a surface in that configuration space. Moreover, this surface either contains all the trajectories in the phase-space, if $Q_0=0$, or it will be asymptotically intersected by the evolution of the system if the scale factor diverges.
  	
  	In the case of a trivial shift-charge, no explicit scale factor dependence is present in the shift-current (\ref{eq:Jsym}). Consequently, the Friedmann (\ref{FE1}) and Raychaudhuri (\ref{FE2}) equations simplify as the term $\dot{\phi}J$ drops out. Then, the evolution of the system could tend to a (quasi)dS state provided that the k-essence function $K$ converges asymptotically to a negative constant and the slow-roll condition $XG_X\ddot{\phi}\approx0$ is satisfied. The presence of a dS future attractor in the shift-symmetric KGB models was first discussed in reference \cite{KGB1} and has indeed attracted considerable attention ever since; see, for instance, references \cite{DeFelice:2010pv,Bernardo:2021hrz,Germani:2017pwt,Martin-Moruno:2015bda,Martin-Moruno:2015kaa,DeFelice:2011bh,Tsujikawa:2010zza}. (These trajectories within the configuration space that lead to a future dS state are sometimes dubbed \textit{tracker} trajectories \cite{Germani:2017pwt,DeFelice:2010pv}.)
  	Conversely, if $Q_0\neq0$, the shift-current is not exactly zero but scales with the expansion. As a result, this scenario may present a broader phenomenology than in the previous case. In fact, if the scalar field velocity increases faster than $a^3$ with the expansion, then the contribution of $\dot{\phi}J$ to the total energy and pressure may diverge. This could lead the evolution of the model towards a different future fate from that of an asymptotic dS state. Moreover, the energy density and pressure of the scalar field may diverge if $\dot{\phi}J$ blows-up; see equations (\ref{eq:rhoX}) and (\ref{eq:PressureX}). Therefore, the future evolution in that case could entail a BR singularity provided that the divergence takes place at a finite cosmic time. This possibility was briefly discussed in our previous work \cite{TMPletter}. (The future phenomenology for a non-trivial shift-charge was also explored in references \cite{Helpin:2019kcq,Muharlyamov:2021dlh,Muharlyamov:2021ixt}.)
    	
  	The existence of future cosmological singularities in these shift-symmetric models is properly addressed in the next section with a dynamical system formulation of the KGB theory. This allows for a systematic study of the fixed points of the theory and their stability.
  	
  	
  	\section{Autonomous system\label{sec:Dynamical Systems}}
 	 	
 	In view of the Friedmann equation (\ref{FE1}), we define the dimensionless variables
 	\begin{align}
 		\Omega_r&\coloneqq \frac{\rho_r}{3H^2},\label{varOR}\\
 		\Omega_m&\coloneqq \frac{\rho_m}{3H^2},\\
 		\Omega_\phi&\coloneqq \frac{\epsilon\sqrt{2X} J-K}{3H^2}\label{varOP},
 	\end{align}
 	where $\epsilon\coloneqq \textup{sgn}\ \dot{\phi}$	labels the increasing and decreasing branches for the scalar field. 
	Moreover, we assume $\Omega_\phi$ to be positive since we are mainly interested in the future attractors of expanding FLRW models. Hence, $\Omega_i\in[0,1]$ for $i\in\lbrace r, m, \phi\rbrace$.
	In terms of these variables, the Friedmann equation (\ref{FE1}) can be expressed as
 	\begin{eqnarray}\label{eq:FriedConstraint}
 		\Omega_r+\Omega_m+\Omega_\phi=1.
 	\end{eqnarray}
 	This relation can be used to eliminate one of the aforementioned variables from the dynamical system. Then, a new independent variable should be introduced in order to obtain an autonomous system. We select this new variable, $h$, as the following compactification scheme for the Hubble rate \cite{FPatInfty} 	
 	\begin{align}\label{varh}
 		\frac{H}{H_0}=\frac{h}{1-h^2},
 	\end{align}
 	being $H_0$ the current value of the Hubble parameter. Note that this transformation\footnote{This compactification is similar to the \textit{arctan} (or \textit{arctanh}) prescription. However, the polynomial compactification (\ref{varh}) was argued to be more convenient for the proper identification and classification of the fixed points, if any, at $H$-infinity \cite{FPatInfty}.} represents a bijective mapping of the $H$-line onto the compact $[-1,1]$. However, since we are not interested in contracting FLRW models ($H<0$), we shall restrict to $h\in[0,1]$. It is also important to highlight that the use of compact variables is strongly recommended, otherwise fixed points at the infinite boundary of the system, if any, may be overlooked.
 	In terms of these new variables the evolution of the system given by equations (\ref{FE2}), (\ref{eq:conservationRad}) and (\ref{eq:phi}) reads
 	\begin{align}
 		h'&=\frac{(1-h^2)h}{1+h^2}C_1,\label{eq:Ph}\\
 		\Omega_r'&=-2\Omega_r\left(2+C_1\right),\label{eq:POR}\\
 		\Omega_\phi'&=C_2-2\Omega_\phi C_1,\label{eq:POP}
 	\end{align}
 	with the auxiliary functions	
 	\begin{align}
 		&C_1\coloneqq\frac{H'}{H},\label{def:C1}\\
 		&C_2\coloneqq\frac{\epsilon\sqrt{2X}}{H^2}\left(H G_XX'-{J}\right),\label{def:C2}
 	\end{align}
 	which, in general, depend on the variables $h$, $\Omega_\phi$ and $\Omega_r$ since $H'$ and $X'$ can be re-expressed in terms of these variables; please find the details in appendix \ref{app:functions}. [Recall that $\Omega_m$ has been eliminated from the dynamical system by means of the Friedmann constraint (\ref{eq:FriedConstraint}).] The prime in the above expressions denotes differentiation with respect to the dimensionless time-like variable $x\coloneqq\ln(a/a_0)$. It should be emphasised, however, that this definition for the independent variable, $x$, of the system is only well-defined for monotonically expanding geometries.   Therefore, recollapsing (turnaround) cosmologies or bounce-like events are excluded from our analysis. In fact, the dynamical variables (\ref{varOR})-(\ref{varOP}) are not even well-suited for addressing the existence of fixed points corresponding to these events. Since a bounce/turnaround would take place at a finite scale factor with vanishing Hubble rate the partial densities (\ref{varOR})-(\ref{varOP}) we have selected as the dynamical variables would diverge. (Recollapsing cosmologies and bounce solutions in KGB theories have been previously addressed, for instance, in reference \cite{Helpin:2019kcq}.)
 	Equilibrium points corresponding to purely contracting FLRW models are also excluded from our discussion.
 	Nevertheless, it should be pointed out that each fixed point for the expanding geometry would have an exact counterpart in a contracting universe due to the symmetry of the background equations under inversion of time. The stability in the contracting regime would be the opposite to that of the expanding case since reversing the time (i.e. $x\to-x$) also reverses the flow defined by (\ref{eq:Ph})-(\ref{eq:POP}) and, therefore, the stability of the equilibrium points. (The reader may find further information on dynamical systems and their applications to cosmology in references \cite{BookColey,BookPerko,BookWainwirght,Boehmer:2014vea,Bahamonde:2017ize}.)	
  	
   	The auxiliary functions $C_1$ and $C_2$ can be seen as functions on the new variables defined in (\ref{varOR}), (\ref{varOP}) and (\ref{varh}). That is $C_i(h,\Omega_\phi,\Omega_r)$ for $i\in\{1,2\}$. These functions are connected with the effective equation of state parameter of the total fluid and the equation of state parameter of the scalar field contribution. The former is
  	\begin{align}
  		w_\textup{eff}&\coloneqq\frac{P_{\textup{tot}}}{\rho_{\textup{tot}}}= 		-1-\frac23C_1.
  	\end{align}
  	Whereas for the latter, one can directly read from the field equations (\ref{FE1}) and (\ref{FE2}) that
  	\begin{align}
  		\rho_\phi&\coloneqq \epsilon\sqrt{2X} J-K,\label{eq:rhoX}\\
  		P_\phi&\coloneqq K-\epsilon\sqrt{2X}H G_X X',\label{eq:PressureX}
  	\end{align}
  	and, therefore,
  	\begin{align}\label{eq:wPhi}
  		w_\phi\coloneqq&\frac{P_\phi}{\rho_\phi}=-1-\frac{1}{3\Omega_\phi}C_2.
  	\end{align}
  	Hence, the auxiliary functions $C_1$ and $C_2$ useful for the realisation of a closed dynamical system are also of physical interest when characterizing the fixed points of the system. 
 	
 	Owing to the general structure of the dynamical equations (\ref{eq:Ph})-(\ref{eq:POP}), the fixed points $(h^{\textup{fp}},\Omega_\phi^{\textup{fp}},\Omega_r^{\textup{fp}})$  of the system can be divided into five different groups\footnote{Recall that the physical interpretation of these groups is from the point of view of expanding FLRW only. Fixed points corresponding to recollapsing (turnaround) cosmologies or bounce-like events cannot be addressed within our formulation. In addition, there may be fixed points in a expanding FLRW universe that have eluded this classification due to the choice of the dynamical variables being not adequate for them to be properly identified. That may be the case when the auxiliary functions $C_1$ and $C_2$ diverge within the phase-space. A way around this issue is discussed in Section \ref{sec:PureBraiding}.}, where the superscript ``fp'' denotes the value of the corresponding quantity at the fixed point. These groups are defined as follows, where $C_i^{\textup{fp}}$ should be read as $C_i(h^{\textup{fp}},\Omega_\phi^{\textup{fp}},\Omega_r^{\textup{fp}})$:
	
 		\paragraph{Group 1}
 		$(h^{\textup{fp}}=\Omega_r^{\textup{fp}}=0$, $C_1^{\textup{fp}}\neq-2$ and $C_2^{\textup{fp}}=2C_1^{\textup{fp} }\Omega_\phi^{\textup{fp}}$): 
 		Vacuum solutions. The evolution of the system in the neighbourhood of these fixed points is either dominated by matter or the scalar field.
 		
 		\paragraph{Group 2}
 		($h^{\textup{fp}}=0$, $C_1^{\textup{fp}}=-2$ and $C_2^{\textup{fp}}=-4\Omega_\phi^{\textup{fp}}$):
 		Vacuum solutions where radiation like effects dominates the nearby evolution of the system, i.e. $w_{\textup{eff}}^{\textup{fp}}=1/3$. Scaling solutions for the scalar field. 
 		
 		\paragraph{Group 3}
 		($h^{\textup{fp}}=1$, $\Omega_r^{\textup{fp}}=0$, $C_1^{\textup{fp}}\neq-2$ and $C_2^{\textup{fp}}=2C_1^{\textup{fp}} \Omega_\phi^{\textup{fp}}$):
 		Cosmological singularities where $H$ and its cosmic time derivative diverge but $\dot{H}/H^2$ (that is $C_1$) remains finite; e.g. BR singularity.
 		
 		\paragraph{Group 4}
 		($h^{\textup{fp}}=1$, $C_1^{\textup{fp}}=-2$ and $C_2^{\textup{fp}}=-4\Omega_\phi^{\textup{fp}}$):
 		Initial cosmological singularities where $H$ and its cosmic time derivative diverge and scaling solutions for the scalar field. It is a radiation dominated regime ($w_{\textup{eff}}^{\textup{fp}}=1/3$). E.g. radiation-induced Big Bang (BB) singularity.
 		
 		\paragraph{Group 5}
 		($h^{\textup{fp}}\neq\{0,1\}$, $\Omega_r^{\textup{fp}}=0$ and $C_1^{\textup{fp}}=C_2^{\textup{fp}}=0$):
 		These fixed points necessary obey $\Omega_m^{\textup{fp}}=0$ since $w_{\textup{eff}}^{\textup{fp}}=-1$. Hence, the scalar field is dominant ($\Omega_\phi^{\textup{fp}}=1$). Moreover, $\Omega_r^{\textup{fp}}=\Omega_m^{\textup{fp}}=0$ and $h^{\textup{fp}}\in(0,1)$ imply, in general, that $a^{\textup{fp}}\to\infty$. These may represent the asymptotic de Sitter solutions of the theory.
 	
 	The existence and stability of the above discussed fixed points depend, ultimately, on the choice for the functions $K$ and $G$. The study of the unstable (repellers) and stable (attractors) equilibrium points is a useful approach to the cosmological evolution of the model since trajectories in the phase-space are known to evolve from the former to the latter equilibrium points. Nevertheless, the dynamical system approach only provides qualitative information of the solution to the background equations (\ref{FE1}), (\ref{FE2}), (\ref{eq:conservationDust}), (\ref{eq:conservationRad}) and (\ref{eq:phi}). This information must be combined with a close inspection of the background equations themselves to obtain as much information as possible on the whole evolution of the system.

 	Once the expressions for the functions $K$ and $G$ are specified, the background equations (\ref{FE1}), (\ref{FE2}) and (\ref{eq:phi}) lead to the auxiliary functions $C_1$ and $C_2$, which, fully characterize the fixed points discussed above; see appendix \ref{app:functions}. However, definition (\ref{varOP}) must be inverted for $X=X(h,\Omega_\phi)$ in order to express the autonomous system in the new variables $(h,\Omega_\phi,\Omega_r)$. This suppose the main limiting factor of our approach as that inversion may not always be possible analytically. In the following sections we present some simple but enlightening examples where this inversion is unambiguous.  
 	
	
	\section{Limiting power law models\label{sec:powerLaws}}

	For the sake of simplicity in the discussion of the future phenomenology of an expanding FLRW universe in KGB theories we consider a power law for the functions $K$ and $G$. That is \cite{DeFelice:2011bh} (see also, for example, references \cite{DeFelice:2011aa,DeFelice:2011th,Giacomello:2018jfi,Frusciante:2019puu})
	\begin{align}
		K(X)={c}_K X^\alpha\hspace{0.5cm} \textup{and} \hspace{0.5cm}G(X)={c}_G X^\beta,
	\end{align}
	being $c_K$ and $c_G$ coupling constants, and $\alpha$ and $\beta$ the parameters labelling different models. 
	The definition (\ref{varOP}), then, reduces to
	\begin{eqnarray}\label{eq:XonOPh}
	3H^2\Omega_\phi=(2\alpha-1)c_K X^{\alpha}+6\sqrt{2}\epsilon c_G\beta  H  X^{\beta+\frac12}.
	\end{eqnarray}
	Recall that this expression must be inverted for $X=X(h,\Omega_\phi)$, taking also into account the definition (\ref{varh}), in order to obtain the closed dynamical system (\ref{eq:Ph})-(\ref{eq:POP}). The limiting models when only the $k$-essence function $K$ or the braiding function $G$ are present are discussed below. A proxy example where both functions are not null is analysed in section \ref{sec:Mod3}.
	
	
	\subsection{Kinetic $k$-essence \label{sec:Kessence}}
	
	We first apply the dynamical systems prescription (\ref{eq:Ph})-(\ref{eq:POP}) to the well-known (power law) kinetic $k$-essence \cite{Chiba:1999ka,Scherrer:2004au,Armendariz-Picon:1999hyi,Armendariz-Picon:2000nqq,dePutter:2007ny} subfamily of the action (\ref{eq:actionKGB}). This is given by
	\begin{align}\label{model:pureK}
		K(X)=c_K X^{\alpha} \hspace{0.5cm} \textup{and} \hspace{0.5cm} G(X)=c_G,
	\end{align}
	being $c_K$ and $c_G$ constants. Note that $G=$const gives rise to a boundary term in the action (\ref{eq:actionKGB}) and, therefore, does not contribute to the field equations.

	The shift-current (\ref{eq:J}) for this model is
	\begin{eqnarray}\label{eq:JsymK}
		J=\sqrt{2}\alpha \epsilon c_K X^{\alpha-\frac12}.
	\end{eqnarray}
	Comparing the preceding expression with equation (\ref{eq:Jsym}), it follows that $\alpha\epsilon c_K$ and $Q_0$ should have the same sign. Consequently, the parameter $\epsilon$ (sgn $\dot{\phi}$) is not allowed to change throughout the evolution of the system. Next, the energy density of the scalar field reads
	\begin{align}\label{eq:rhoMod1}
		\rho_\phi=(2\alpha-1)c_K X^\alpha.
	\end{align}
	Please note that demanding this energy density to be positive throughout the evolution results in the constraint $(2\alpha-1)c_K>0$. 
	
	The inversion of equation (\ref{eq:XonOPh}) for the models at hands leads to
	\begin{eqnarray}\label{eq:invOPpureK}
		X=\left[\frac{3H_0^2h^2\Omega_\phi}{c_K(2\alpha-1) (1-h^2)^2}\right]^\frac{1}{\alpha},
	\end{eqnarray}
	where the quantity in brackets is always positive.
	Then, the functions $C_1$ and $C_2$ read
	\begin{align}
		C_1(\Omega_\phi,\Omega_r)&=-\frac12\left(3+\Omega_r+\frac{3\Omega_\phi}{2\alpha-1}\right),\label{eq:C1K}\\
		C_2(\Omega_\phi)&=-\frac{6\alpha \Omega_\phi}{2\alpha-1},\label{eq:C2K}
	\end{align}
	see definitions in equations (\ref{def:C1}) and (\ref{def:C2}), respectively. It should be noted that these expressions do not depend explicitly on $h$, $c_K$ or $\epsilon$ since they have been completely absorbed into the partial densities $\Omega_i$. In addition, the function $C_2$ depends only on the scalar field partial density. That is to be expected as equation (\ref{eq:wPhi}) depends only on the kinetic term $X$ when braiding term is absent.
	
	The fixed points of this model with their stability and physical interpretation are shown in table \ref{tab:fpkessence}.
	\begin{table*}
		\begin{center}
			\setlength\arrayrulewidth{0.6pt}
			\renewcommand{\arraystretch}{1.5}
			\begin{tabular}{l c c c c c c c c c}
				\hline \hline 
				Fixed Point & $(h^{\textup{fp}},\Omega_\phi^{\textup{fp}},\Omega_r^{\textup{fp}})$ & $w_X^{\textup{fp}}$ & $w_{\textup{eff}}^{\textup{fp}}$ & $\alpha<0$ & $\alpha=0$ & $0<\alpha<\frac12$ &  $\frac12<\alpha<2$ &
				$\alpha=2$ & $2<\alpha$ \\ \hline
				A$_1$ (vacuum) & $(0,0,0)$ & $\frac{1}{2\alpha-1}$ & $0$ & saddle & saddle & saddle &  attractor &  attractor & attractor \\ 
				B$_1$ (vacuum) & $(0,1,0)$ &  $\frac{1}{2\alpha-1}$ &  $\frac{1}{2\alpha-1}$ & attractor & | & saddle  & saddle & | & saddle \\ 
				C$_1$ (vacuum) & $(0,0,1)$ & $\frac{1}{2\alpha-1}$ & $\frac{1}{3}$ & saddle & saddle & saddle &  saddle & | & saddle  \\ 
				D$_1$ (BB)& $(1,0,0)$ & $\frac{1}{2\alpha-1}$ &  $0$ & saddle & saddle & saddle & saddle & saddle & saddle \\ 
				E$_1$ (BB/BR)& $(1,1,0)$ &  $\frac{1}{2\alpha-1}$ &   $\frac{1}{2\alpha-1}$ & saddle & | & attractor & repeller & | & saddle \\ 
				F$_1$ (BB)& $(1,0,1)$ &  $\frac{1}{2\alpha-1}$ & $\frac{1}{3}$ & repeller & repeller & repeller &  saddle & | & repeller \\ 
				$\mathcal{A}_1$ (vacuum)& $(0,\Omega_\phi^{\textup{fp}},\Omega_r^{\textup{fp}})$ & $\frac{1}{3}$ & 	$\frac{1}{3}$ & | & | & | &  | & saddle & | \\ 
				$\mathcal{B}_1$ (BB)& $(1,\Omega_\phi^{\textup{fp}},\Omega_r^{\textup{fp}})$ & $\frac{1}{3}$ & $\frac{1}{3}$ & | & | & | & | & repeller & | \\ 
				$\mathcal{S}_1$ (dS)& $(h^{\textup{fp}},1,0)$ & $-1$ & $-1$ & | & attractor & | & | & | & | \\ 
				\hline 
			\end{tabular}
		\end{center}
	\caption{Classification and linear stability of the fixed points of the model (\ref{model:pureK}). 	A superscript ``fp'' denotes evaluation at the fixed point whereas a horizontal bar indicates that the corresponding fixed point does not exist. The calligraphic characters $\mathcal{A}_1$, $\mathcal{B}_1$ and $\mathcal{S}_1$ label normally hyperbolic equilibrium sets. The condition $\Omega_\phi^{\textup{fp}}+\Omega_r^{\textup{fp}}=1$ holds for  $\mathcal{A}_1$ and $\mathcal{B}_1$. In addition, $h^{\textup{fp}}$ may take any values (different form 0 or 1) in $\mathcal{S}_1$.\label{tab:fpkessence} }
	\end{table*}
	The points from A$_1$ down to F$_1$ are hyperbolic equilibrium points and, therefore, their stability follows from the usual linear theory. Conversely, $\mathcal{A}_1$, $\mathcal{B}_1$ and $\mathcal{S}_1$ conform three sets of non-isolated non-hyperbolic fixed points. For each of these equilibrium sets, one of the eigenvalues of the Jacobian matrix is zero. However, the null eigenvalue corresponds to the eigenvector tangent to the set containing the non-isolated equilibrium points. These equilibrium sets are said to be normally hyperbolic \cite{BookColey,BookWainwirght} and their stability is given by the real part of the eigenvalues in the remaining directions. 
	
	Note that there is only one attractor and one repeller in the configuration space for a given value of the parameter $\alpha$; see table \ref{tab:fpkessence}. Physical trajectories in the phase-space will start at the corresponding repeller and will univocally evolve towards the attractor equilibrium point, maybe passing close to a saddle point. It should be emphasised, however, that the classification provided in table \ref{tab:fpkessence} contains only qualitative information of the would-be complete solution to background equations (\ref{FE1}), (\ref{FE2}), (\ref{eq:conservationDust}), (\ref{eq:conservationRad}) and (\ref{eq:phi}). This information should be combined with a close inspection of the background equations themselves to obtain as much information as possible on the particular dynamics of each trajectory.
	
	Recall that we have focused our analysis on expanding geometries only and, therefore, the physical interpretation of the points in table \ref{tab:fpkessence} is deduced according to that ansatz.
	Equilibrium points corresponding to bounce or turnaround-like events, if any, cannot be described within this approach. (However, each of the points in table \ref{tab:fpkessence} would have an exact counterpart in a monotonically contracting cosmos with precisely the opposite stability and where $h^{\textup{fp}}\to-h^{\textup{fp}}$ should be applied.)
	According to this interpretation, and taking also into account the background equations (\ref{FE1}), (\ref{FE2}), (\ref{eq:conservationDust}), (\ref{eq:conservationRad}) and (\ref{eq:phi}), it follows that the equilibrium points A$_1$, B$_1$ and C$_1$, and the equilibrium set $\mathcal{A}_1$ correspond to vacuum solutions where all the components of the universe are, eventually, redshifted away with the expansion. Moreover, $\mathcal{A}_1$ contains scaling solutions where the scalar field mimics radiation. The points A$_1$ and B$_1$ belong to group 1 in our previous classification. Conversely, C$_1$ and $\mathcal{A}_1$ are part of group 2. 
	
	The fixed point D$_1$ may be interpreted as an initial matter-induced BB singularity where only some trajectories (those where radiation is absent in the early universe) may begin at D$_1$ if matter dominates over the scalar field in the asymptotic past of the system. However, since a non-trivial radiation content will always dominate over matter at early time, D$_1$ necessary acts as a saddle point in the configuration space. This point belongs to group 3.

	At the equilibrium point E$_1$ the scalar field drives the divergence of both $H$ and $\dot{H}$. This may have different physical interpretations according to the value of $\alpha$. Since the scalar field is the dominant component at E$_1$, then the approximation
	\begin{eqnarray}\label{eq:approxFE1 kessence}
		3H^2\approx \lambda_1 \left(\frac{a}{a_0}\right)^{-\frac{6\alpha}{2\alpha-1}},
	\end{eqnarray}
	holds true, where $\lambda_1\coloneqq(2\alpha-1)c_K\left({Q_0/\sqrt{2}\epsilon\alpha c_K}\right)^\frac{2\alpha}{2\alpha-1}$ is a positive constant. [Recall that $Q_0$ and $\epsilon \alpha c_K$ have the same sign. In addition, $(2\alpha-1)c_K$ is positive in an expanding universe.] 
	For $\alpha\in(1/2,2)$ the scalar field dominates over radiation in the very early universe, thus, leading to the divergence of the Hubble rate and its time derivative as $a\to0$. Hence, a scalar-field-induced BB singularity takes place. On the other hand, if $\alpha\in(0,1/2)$, the Hubble rate becomes proportional to a positive power of the scale factor. It is a well-known result that in this situation $a$, $H$ and $\dot{H}$ blow-up in a finite cosmic time (see appendix \ref{app:BR} for a justification of this claim). Consequently, the model (\ref{model:pureK}) entails a BR singularity when  $\alpha\in(0,1/2)$. In fact, this is the only future attractor in the configuration space for that set of values of $\alpha$; see table \ref{tab:fpkessence}. For the rest of the $\alpha$-line, the exponent in equation (\ref{eq:approxFE1 kessence}) is negative but greater than -4 (radiation). This leads to saddle configurations that can be interpreted in the same fashion as for D$_1$. The equilibrium point E$_1$ belong to group 3 in the discussion of the previous section.
	
	F$_1$ represents a radiation dominated BB singularity. It naturally acts as a repeller in the configuration space except for those value of $\alpha$ for which the scalar field dominates at the very early universe. An interesting subcase of this event is when $\alpha=2$, in this scenario the scalar field scales exactly as radiation. This scaling solution also corresponds to a radiation-induced BB singularity; see $\mathcal{B}_1$ in table \ref{tab:fpkessence}. Both of these scenarios belong to group 4 in our previous classification.
	
	It should be noted that $\alpha=0$ corresponds to the standard $\Lambda$CDM model where the role of the cosmological constant is portrayed by the coupling constant $c_K$; that is $\Lambda=-c_K$ where $c_K<0$ (since $\rho_\phi$ positive). Therefore, the expansion history of the model for $\alpha=0$ would be that of $\Lambda$CDM. 	That is, the system would evolve towards a future dS state\footnote{This is indeed the only de Sitter solution for the power law kinetic k-essence model at hands.} provided that $c_K$ is not null (see $\mathcal{S}_1$ in table \ref{tab:fpkessence}). The scalar field dominated fixed points B$_1$ and E$_1$ are not present in this case since they would correspond to $c_K=0$ (i.e. no scalar field) and $c_K\to\infty$ (unphysical), respectively. 
	
	Finally, it should be also mentioned that for $\alpha=1/2$ the scalar field energy density (\ref{eq:rhoMod1}) is trivial. Hence, the corresponding universe is filled with dust and radiation only. This scenario is not included in table \ref{tab:fpkessence} as our interest resides mainly in the phenomenology of the scalar field.

	
	\subsection{Pure \textit{braiding} \label{sec:PureBraiding}}
	
	A proxy model exhibiting the interesting phenomenology of the KGB theory is that when only the $G$ function is present. That is  
	\begin{align}\label{model:pureG}
		K(X)=0 \hspace{.5cm}\textup{and}\hspace{0.5cm} G(X)=c_G X^\beta,
	\end{align}
	being $c_G$ a coupling constant and $\beta$ the parameter labelling different models. This is the model briefly considered in reference \cite{TMPletter}. In that case, the shift-current (\ref{eq:J}) reduces to
	\begin{eqnarray}\label{eq:JsymG}
		J=6\beta c_G H X^{\beta}.
	\end{eqnarray}
	Therefore, the energy density of the scalar field reads
	\begin{align}\label{eq:rhoMod2}
		\rho_\phi=6\sqrt{2}\epsilon c_G\beta  H  X^{\beta+\frac12}.
	\end{align}
	Assuming this energy density to be non-negative yields the restriction $\beta\epsilon c_GH\geq0$. In an expanding universe ($H>0$), this condition implies that $\epsilon$ cannot change its sign throughout the evolution of the system.
	Then, from equation (\ref{eq:XonOPh}) it follows that
	\begin{eqnarray}\label{eq:XonOPhSoloG}
		X=\left[\frac{H_0h\Omega_\phi}{2\sqrt{2}\beta\epsilon c_G(1-h^2)}\right]^\frac{2}{2\beta+1},
	\end{eqnarray}
	where the quantity in brackets is always positive.
	
	The auxiliary functions $C_1$ and $C_2$ for this model read
	\begin{align}
		C_1(\Omega_\phi,\Omega_r)&=-\frac{2\beta\Omega_r+6\beta+3\Omega_\phi}{4\beta+\Omega_\phi},\label{eq:C1pureBraid}\\
		C_2(\Omega_\phi,\Omega_r)&=\Omega_\phi\frac{\Omega_r-3-12\beta-3\Omega_\phi}{4\beta+\Omega_\phi},\label{eq:C2pureBraid}
	\end{align}
	see definitions in equations (\ref{def:C1}) and (\ref{def:C2}), respectively. As for the kinetic $k$-essence model, these functions do not explicitly depend on $h$, $c_G$ or $\epsilon$ since they have been completely absorbed in the definitions of the partial densities. However, now the function $C_2$ depends also on $\Omega_r$ due to the \textit{braiding} (recall that this function is related to $w_\phi$ and, therefore, to the evolution of the scalar field).
	
	The fixed points of this model are shown in table \ref{tab:fpBraiding}. The points from A$_2$ down to F$_2$ are hyperbolic equilibrium points of the system and their stability follows from the usual linear theory. The labels $\mathcal{A}_2$, $\mathcal{B}_2$ and $\mathcal{S}_2$ denote normally hyperbolic equilibrium sets. 
	On the other hand, G$_2$ and $\mathcal{C}_2$ in table \ref{tab:fpBraiding} represent events that have eluded our analysis because of the choice for the dynamical variables being not adequate for them to be properly identified as equilibrium configurations of the system. Nevertheless, their existences and stability follows directly from the background equations (a discussion we return to below).

	\begin{table*}\small
			\setlength\arrayrulewidth{0.6pt}
			\renewcommand{\arraystretch}{1.5}
			\begin{adjustwidth}{-1.5cm}{}
			\begin{tabular}{l c c c c c c c c c c c}
				\hline \hline 
				Fixed Point & $(h^{\textup{fp}},\Omega_\phi^{\textup{fp}},\Omega_r^{\textup{fp}})$ & $w_\phi^{\textup{fp}}$ & $w_{\textup{eff}}^{\textup{fp}}$ & $\beta<-\frac12$ & $\beta=-\frac12$ & $-\frac12<\beta<-\frac14$ &
				$\beta=-\frac14$ & $-\frac14<\beta<0$ & $0<\beta<\frac12$ & $\beta=\frac12$ & $\frac12<\beta$  \\\hline
				$\textup{A}_2$ (vacuum)& $(0,0,0)$ & $\frac{1}{4\beta}$ & $0$ & saddle  & saddle & saddle & saddle & saddle & attractor & attractor &  attractor \\ 
				$\textup{B}_2$ (vacuum)& $(0,1,0)$ & $\frac{1}{4\beta+1}$ & $\frac{1}{4\beta+1}$ & attractor  & attractor & saddle & | & saddle & saddle & | & saddle \\ 
				$\textup{C}_2$ (vacuum)& $(0,0,1)$ & $\frac{1}{6\beta}$ & $\frac{1}{3}$ & saddle  & saddle & saddle & saddle & saddle & saddle & | &  saddle \\ 
				$\textup{D}_2$ (BB)& $(1,0,0)$ & $\frac{1}{4\beta}$ & 0 & saddle  & saddle & saddle & saddle & saddle & saddle & saddle & saddle \\ 
				$\textup{E}_2$ (BB/BR)& $(1,1,0)$ & $\frac{1}{4\beta+1}$ &  $\frac{1}{4\beta+1}$ & saddle  & | & attractor & | & repeller & repeller & | & saddle \\ 
				$\textup{F}_2$ (BB)& $(1,0,1)$ & $\frac{1}{6\beta}$ & $\frac{1}{3}$ & repeller  & repeller & repeller & repeller & repeller & saddle & | & repeller \\ 
				$\textup{G}_2$ (BF)& $(1,1,0)$ & $-\infty$ & $-\infty$ & |  & | & | & attractor$^\star$ & | & | & | & | \\
				$\mathcal{A}_2$ (vacuum)& $(0,\Omega_\phi^{\textup{fp}},\Omega_r^{\textup{fp}})$ & $\frac{1}{3}$ & $\frac{1}{3}$ & | & | & | & | & | & | & saddle & | \\ 
				$\mathcal{B}_2$ (BB)& $(1,\Omega_\phi^{\textup{fp}},\Omega_r^{\textup{fp}})$ & $\frac{1}{3}$ & $\frac{1}{3}$ & | & | & | & | & | & | & repeller & | \\
				${\mathcal{C}_2}$ (sudden)& $(h^\textup{fp},-4\beta,\Omega_r^{\textup{fp}})$ & $-\infty$ & $-\infty$ & | & | & | & |  &  attractor$^\star$  & | & |  & |  \\ 
				${\mathcal{S}_2}$ (dS)& $(h^\textup{fp},1,0)$ & $-1$ & $-1$ & | & attractor & | & |  & |  & |  & |  & |  \\
				\hline 
			\end{tabular}
			\end{adjustwidth}
			\caption{Classification and linear stability of the fixed points of the model (\ref{model:pureG}). A superscript ``fp'' denotes evaluation at the fixed point whereas a horizontal bar indicates that the corresponding fixed point does not exist. The calligraphic characters $\mathcal{A}_2$, $\mathcal{B}_2$, $\mathcal{C}_2$ and $\mathcal{S}_2$ label sets of non-isolated fixed points where $h^{\textup{fp}}$ may take any values (different from 0 or 1). In addition,   $\Omega_\phi^{\textup{fp}}+\Omega_r^{\textup{fp}}=1$ holds for $\mathcal{A}_2$  and $\mathcal{B}_2$, and $\Omega_r^{\textup{fp}}\in[0,1+4\beta]$ for $\mathcal{C}_2$. The starred quantities indicate fixed points that have eluded our dynamical system analysis because of the choice of the dynamical variables but whose existence and stability follows directly from the background equations.\label{tab:fpBraiding} }
	\end{table*}

	The physical interpretation of the fixed points in table \ref{tab:fpBraiding} is analogous to that of the kinetic $k$-essence scenario discussed before. [Recall that the dynamical system analysis provides only qualitative information that must be combined with a close inspection of the background equations (\ref{FE1}), (\ref{FE2}), (\ref{eq:conservationDust}), (\ref{eq:conservationRad}) and (\ref{eq:phi}) to obtain as much information as possible on the particular dynamics of each trajectory in the phase-space.] Thus, A$_2$, B$_2$, C$_2$ and $\mathcal{A}_2$ represent vacuum solutions in an expanding universe (where A$_2$ and B$_2$ belong to group 1, and C$_2$ and $\mathcal{A}_2$ to group 2). D$_2$ (group 3) denotes a saddle configuration where a matter-induced BB takes place. A radiation-induced BB corresponds to F$_2$ and $\mathcal{B}_2$, where for the latter the scalar field scales exactly as radiation. (Both  these fixed points belong to group 4.)
	
	As in the previous section, E$_2$ (group 3) may have different interpretations depending on the parameter $\beta$. Since the scalar field dominates over dust and radiation, and taking also into account equation (\ref{eq:Jsym}), the Friedmann equation (\ref{FE1}) in an expanding universe ($H>0$) reduces to
	\begin{eqnarray}\label{eq:HapproxE2}
		H\approx\lambda_2 \left(\frac{a}{a_0}\right)^{-3\frac{2\beta+1}{4\beta+1}},
	\end{eqnarray} 
	being $\lambda_2\coloneqq(\sqrt{2}\epsilon Q_0/3)^\frac{2\beta}{4\beta+1}\left(Q_0/6\beta c_G\right)^{\frac{1}{4\beta+1}}$ a positive constant since $\epsilon Q_0$ and $Q_0 \beta c_G$ are positive. [Recall that the former constraint comes from demanding $\rho_\phi$ to be positive whereas the latter follows from comparing equations (\ref{eq:Jsym}) and (\ref{eq:JsymG}).] When $\beta\in(-1/4,1/2)$, the scalar field dominates over radiation as $a\to0$. This results in a scalar-field-induced BB. On the other hand, if $\beta\in(-1/2,-1/4)$ the exponent becomes positive and, therefore, a future BR singularity takes place; see appendix \ref{app:BR}. For the rest of the $\beta$-line, E$_2$ corresponds to saddle configurations where $\rho_\phi$ either dominates or not over matter when $a\to0$.
	
	The case of $\beta=0$ is not portrayed in table \ref{tab:fpBraiding} since $G=$const gives rise to a boundary term in the action (\ref{eq:actionKGB}) and, therefore, the corresponding model would contain dust and radiation only (recall that $K=0$ for the model at hands). Another critical value for $\beta$ is that of $-1/2$. In that case, the energy density of the scalar field depends only on the Hubble rate; see equation (\ref{eq:rhoMod2}). Thus, as matter and radiation are redshifted away the Hubble rate converges to a constant value given by $H=-\sqrt{2}\epsilon c_G$, where $\epsilon c_G\leq0$; cf. $\beta=-1/2$ in equation (\ref{eq:HapproxE2}). This solution corresponds to the dS fixed point of the system (group 5) if $\epsilon$ is not null; see $\mathcal{S}_2$ in table \ref{tab:fpBraiding}. The vacuum equilibrium point B$_2$ (group 1) is obtained if $\epsilon=0$ in the future. Please note that B$_2$ and $\mathcal{S}_2$ lined-up in a set of normally hyperbolic fixed points. We have represented them separately in table \ref{tab:fpBraiding} when $\beta=-1/2$ only to facilitate their physical interpretation. Therefore, for this value of $\beta$ all trajectories in the phase-space will evolve from F$_2$ to the B$_2$-$\mathcal{S}_2$ equilibrium line. Also note that the scalar field dominated fixed point E$_2$ is not present for $\beta=-1/2$ in an expanding universe since it would correspond to $c_G\to\infty$ (unphysical).
	
	Note that special attention should be paid to $\Omega_\phi=-4\beta$, moment at which the  denominator in equations (\ref{eq:C1pureBraid}) and (\ref{eq:C2pureBraid}) cancels. This takes place in the physical phase-space whenever $\beta\in[-1/4,0)$. In that case the dynamical system portrayed by the auxiliary functions $C_1$ and $C_2$ is potentially ill-defined and, therefore, fixed points corresponding to this value for $\Omega_\phi$, if any, would be hardly studied within this formulation. Nevertheless, the behaviour of the model at this situation can be directly inferred from the Friedmann and Raychaudhuri equations. Consider first $\beta=-1/4$, which corresponds to the dynamical system (\ref{eq:Ph})-(\ref{eq:POP}) being potentially ill-defined at $\Omega_\phi=1$. For this value of the exponent $\beta$, the Friedmann equation (\ref{FE1}) reduces to
	\begin{align}
		\left[1-\Omega_{\phi0}\left(\frac{a}{a_0}\right)^3\right]\frac{H^2}{H_0^2}&=\Omega_{r0}\left(\frac{a}{a_0}\right)^{-4}+\Omega_{m0}\left(\frac{a}{a_0}\right)^{-3},
	\end{align}	
	being $\Omega_{r0}$, $\Omega_{m0}$ and $\Omega_{\phi0}$ the present value of the partial densities for radiation, matter and the scalar field, respectively. For an expanding geometry, the expression in brackets on the l.h.s. vanishes at a finite scale factor, namely $a_{\textup{s}}\coloneqq a_0\Omega_{X0}^{-1/3}$. Nevertheless, since the r.h.s. of the preceding equation is different from zero whenever the scale factor $a$ is finite, then, the Hubble rate necessarily diverge when the bracket vanishes. 
	Similarly, the  Raychaudhuri equation (\ref{FE2}) simplifies to
	\begin{align}
		&\left[1-\Omega_{\phi0}\left(\frac{a}{a_0}\right)^3\right]\frac{\dot{H}}{H_0^2}= \frac32\Omega_{\phi0}\left(\frac{a}{a_0}\right)^3\frac{H^2}{H_0^2}-\frac12 \left[4\Omega_{r0}\left(\frac{a}{a_0}\right)^{-4}+3\Omega_{m0}\left(\frac{a}{a_0}\right)^{-3}\right],
	\end{align}
	which implies that  $\dot{H}$ also diverge when the observable universe reaches the maximum size $a_{\textup{s}}$. Moreover, both $H$ and $\dot{H}$ diverge at a finite cosmic time\footnote{This follows from the fact that $1/aH$ is always bounded for $a\in[0,a_{\textup{s}}]$. Hence, $\int_0^{a_\textup{s}}\textup{d}a/aH=\int_0^{t_\textup{s}}\textup{d}t$ is finite.}.	In addition, the scalar field exhibits strong phantom behaviour near $a_{\textup{s}}$. The equation of state parameters $w_{\textup{eff}}$ and $w_X$ even diverge to minus infinity when $H$ and $\dot{H}$ explode. This behaviour corresponds to a BF singularity; see references \cite{BF1BouhmadiLopez:2006fu,BF2BouhmadiLopez:2007qb} and, for instance, the type III singularities in the classification of reference \cite{NojiriClassification}. In fact, this BF singularity takes place for an expanding universe regardless the choice of the initial partial densities. Therefore, this cosmic singularity acts as a genuine attractor in the theory even though the characterization of the system by means of the dynamical variables $h$, $\Omega_\phi$ and $\Omega_r$ is not well-suited for describing this event. For the sake of completeness, this attractor has been added to table \ref{tab:fpBraiding} under the label G$_2$.
	
	A similar line of reasoning with the Friedmann and Raychaudhuri equations,
	\begin{align}
	\frac{H^2}{H_0^2}=&\Omega_{r0}\left(\frac{a}{a_0}\right)^{-4}+\Omega_{m0}\left(\frac{a}{a_0}\right)^{-3}+{\Omega}_{\phi0}\left(\frac{a}{a_0}\right)^{-3\frac{2\beta+1}{2\beta}}\left(\frac{H}{H_0}\right)^{-\frac{1}{2\beta}},
	\end{align}
	\begin{align}
	\left[1+\frac{\Omega_{\phi0}}{4\beta}\left(\frac{a}{a_0}\right)^{-3\frac{2\beta+1}{2\beta}}\left(\frac{H}{H_0}\right)^{-\frac{4\beta+1}{2\beta}}\right]\frac{\dot{H}}{H_0^2}=&-2\Omega_{r0}\left(\frac{a}{a_0}\right)^{-4}-\frac32\Omega_{m0}\left(\frac{a}{a_0}\right)^{-3}\nonumber\\
	&-\frac{3\Omega_{\phi0}(2\beta+1)}{4\beta}\left(\frac{a}{a_0}\right)^{-3\frac{2\beta+1}{2\beta}} \left(\frac{H}{H_0}\right)^{-\frac{1}{2\beta}},
	\end{align}
	respectively, concludes that $\dot{H}$ diverge at a finite value for $a$ and $H$ whenever $\beta\in(-1/4,0)$. That occurs when the bracket in the l.h.s of the latter equation cancels, which corresponds to $\Omega_\phi=-4\beta$ in terms of our dynamical system variables. Moreover, numerical integrations  for different values for $\beta$, $\Omega_{r0}$, $\Omega_{m0}$ and $\Omega_{\phi0}$ confirm that this cancellation happens at a finite cosmic time. This behaviour corresponds to that of a sudden singularity \cite{sudden1} (see also type II singularities in the classification of reference \cite{NojiriClassification}). Since an expanding system always evolves towards this scenario regardless the choice for the initial partial densities, this event has been added to table \ref{tab:fpBraiding} as an attractor in the corresponding phase-space; see  $\mathcal{C}_2$ in table \ref{tab:fpBraiding}. It should be also mentioned that two repellers are simultaneously present when $\beta\in(-1/4,0)$. This is because $\mathcal{C}_2$ acts as a separatrix dividing the phase-space into two separated parts, where each of the halves contains one of the repellers. The trajectories in each part of the phase-space will begin at the corresponding repeller (E$_2$ or F$_2$) and will evolve towards $\mathcal{C}_2$.
	
	Up-to our knowledge, this is the first time a BR, a BF or a sudden cosmic singularity have been explicitly described in the shift-symmetric sector of the KGB theory. [We also refer the reader to the companion reference \cite{TMPletter} for a discussion on the expansion history of the model (\ref{model:pureG}).] These results suggest that the future phenomenology of the KGB theory may be richer than previously considered. Furthermore, the presence of finite-size singularities illustrates that $J=0$ is not, in general, an exhaustive characterization of all the possible future attractors for an expanding universe in the shift-symmetric KGB theory. This is because the observable universe reaches a maximum size and, therefore, in view of expression (\ref{eq:Jsym}), $J\neq0$ on the future attractor provided that the shift-charge $Q_0$ is non-trivial.
	
	
	\section{Proxy kinetic gravity braiding model\label{sec:Mod3}}
	
	An example featuring both functions $K$ and $G$ is that given by
	\begin{align}\label{model:fullKGB1}
		K(X)=c_K X^{\alpha}\hspace{.5cm}\textup{and}\hspace{0.5cm} 
		G(X)=c_G X^{\alpha-\frac12},
	\end{align}
	being $c_K$ and $c_G$ coupling constants. Note that this is a subclass of the extended Galileon models studied in references \cite{DeFelice:2011bh,DeFelice:2011aa,Giacomello:2018jfi}.	The shift current (\ref{eq:J}), then, reduces to
	\begin{eqnarray}\label{eq:JsymMod3}
		J=\left[\sqrt{2}\alpha \epsilon c_K+3(2\alpha-1)c_G H\right]X^{\alpha-\frac12}.
	\end{eqnarray}
	Accordingly, the energy density of the scalar field reads
	\begin{align}\label{eq:rhoMod3}
		\rho_\phi=(2\alpha-1)\left(c_K+3\sqrt{2}\epsilon c_GH\right)X^\alpha.
	\end{align}
	Demanding this energy density to be positive, at least when the scalar is dominant, is not so straightforward as for the previous models. This is because the parameter $\epsilon$ is not fixed (as it was in the previous examples) and, therefore, it could change its sign throughout the evolution of the system. We explore the necessary conditions for this energy density to be positive, at least when the scalar field is dominant,  in section \ref{sec:positivityED}.
	
	Taking into account the definitions (\ref{varOP}) and (\ref{varh}), equation (\ref{eq:rhoMod3}) yields
	\begin{eqnarray}
		X=\left[\frac{3H^2_0 \gamma h^2 \Omega_\phi}{(2\alpha-1)c_K\left(1-h^2\right)\left(\gamma(1-h^2)+h\right)}\right]^{\frac{1}{\alpha}},
	\end{eqnarray}
	where $\gamma\coloneqq c_K/(3\sqrt{2}\epsilon H_0 c_G)$ is a dimensionless quantity introduced for the sake of the notation.
	The corresponding functions $C_1$ and $C_2$ read
	\begin{align}
		C_1(h,\Omega_\phi,\Omega_r)&=-\frac{\left(2\gamma\alpha(1-h^2) +(2\alpha-1) h\right)\left(\gamma(1-h^2)+h\right)\left((2\alpha-1)(3+\Omega_r)+3\Omega_\phi\right)}{(2\alpha-1)\left[4\alpha\left(\gamma(1-h^2)+h\right)^2-2 h\left(\gamma(1-h^2)+h\right)+h^2\Omega_\phi\right]},\label{eq:C1KGBmod3}\\
		C_2(h,\Omega_\phi,\Omega_r)&=-\Omega_\phi\frac{24\alpha^2\left(\gamma(1-h^2)+h\right)^2-12\alpha h\left(\gamma(1-h^2)+h\right)+(2\alpha-1)h^2\left(3\Omega_\phi-\Omega_r-3\right)}{(2\alpha-1)\left[4\alpha\left(\gamma(1-h^2)+h\right)^2-2 h\left(\gamma(1-h^2)+h\right)+h^2\Omega_\phi\right]},\label{eq:C2KGBmod3}
	\end{align}
	where the interplay between both $K$ and $G$ functions has now introduced an explicit dependence on $h$. In addition, there is also an explicit dependence on the parameter $\gamma$. That should not be surprising as in this scenario there are two coupling constants for the scalar field and, therefore, the dynamics of the system is expected to depend on their ratio.
	\begin{table*}\small
			\setlength\arrayrulewidth{0.6pt}
			\renewcommand{\arraystretch}{1.5}
			\begin{adjustwidth}{-1.8cm}{}
			\begin{tabular}{l c c c c c c c c c c c c}
				\hline \hline 
				Fixed Point & $(h^{\textup{fp}},\Omega_\phi^{\textup{fp}},\Omega_r^{\textup{fp}})$ & $w_X^{\textup{fp}}$ & $w_{\textup{eff}}^{\textup{fp}}$ & $\alpha<0$ & $0<\alpha<\frac14$ & $\alpha=\frac14$ &
				$\frac14<\alpha<\frac12$ & $\frac12<\alpha<1$ & $\alpha=1$ & $1<\alpha<2$ & $\alpha=2$ & $2<\alpha$  \\\hline
				A$_3$ (vacuum) & $(0,0,0)$ & $\frac{1}{2\alpha-1}$ & $0$ & saddle & saddle & saddle & saddle & attractor & attractor & attractor &  attractor & attractor \\ 
				B$_3$ (vacuum) & $(0,1,0)$ & $\frac{1}{2\alpha-1}$ & $\frac{1}{2\alpha-1}$ & attractor & saddle  & saddle & saddle & saddle & saddle & saddle & | & saddle \\ 
				C$_3$ (vacuum) & $(0,0,1)$ & $\frac{1}{2\alpha-1}$ & $\frac{1}{3}$ & saddle & saddle & saddle & saddle & saddle & saddle & saddle &  | &saddle \\ 
				D$_3$ (BB) & $(1,0,0)$ & $\frac{1}{4\alpha-2}$ & 0 & saddle & saddle & saddle & saddle & saddle & saddle & saddle & saddle & saddle \\ 
				E$_3$ (BB/BR) & $(1,1,0)$ & $\frac{1}{4\alpha-1}$ &  $\frac{1}{4\alpha-1}$ & saddle & attractor & | & repeller & repeller & | & saddle & saddle & saddle \\ 
				$\textup{F}_3$ (BB) & $(1,0,1)$ & $\frac{1}{6\alpha-3}$ & $\frac{1}{3}$ & repeller & repeller & repeller & repeller & saddle & | & repeller & repeller & repeller \\
				S$_3^{1}$ (dS) & $(h_{1},1,0)$ & $-1$ & $-1$ & attractor & attractor & attractor & attractor & attractor & attractor & attractor & attractor & attractor \\
				S$_3^{2}$ (dS) & $(h_{2},1,0)$ & $-1$ & $-1$ & saddle &  attractor & attractor & attractor & saddle & saddle & saddle & saddle & saddle \\ 
				$\mathcal{A}_3$ (vacuum) & $(0,\Omega_\phi^{\textup{fp}},\Omega_r^{\textup{fp}})$ & $\frac{1}{3}$ & $\frac{1}{3}$ & | & | & | & | & | & | & | & saddle & | \\ 
				$\mathcal{B}_3$ (BB) & $(1,\Omega_\phi^{\textup{fp}},\Omega_r^{\textup{fp}})$ & $\frac{1}{3}$ & $\frac{1}{3}$ & | & | & | & | & | & repeller & | & | & | \\
				\hline 
			\end{tabular}
			\end{adjustwidth}
			\caption{Classification and linear stability of the fixed points of the model (\ref{model:fullKGB1}). A superscript ``fp'' denotes evaluation at the fixed point whereas a horizontal bar indicates that the corresponding fixed point does not exist. The calligraphic characters $\mathcal{A}_3$ and $\mathcal{B}_3$ label normally hyperbolic equilibrium sets where the condition   $\Omega_\phi^{\textup{fp}}+\Omega_r^{\textup{fp}}=1$ holds. The expressions for $h_1$ and $h_2$ can be found in equations (\ref{eq:h1Mod3}) and (\ref{eq:h2Mod3}), respectively.\label{tab:fpModelKGB3} }
	\end{table*}
	Also note that these expressions for the functions $C_1$ and $C_2$ reduce to those presented in section \ref{sec:Kessence} when $\gamma\to\infty$ (i.e. $c_G\to0$), and to those in section \ref{sec:PureBraiding} when $\gamma\to0$ (that is $c_K\to0$), as to be expected.
	
	
	\subsection{Fixed points\label{sec:fpMod3}}
	
	The fixed point of this model are listed in table \ref{tab:fpModelKGB3}. Following the previous notation, the points form A$_3$ down to S$_3^2$ are hyperbolic equilibrium points whose stability has been deduced linearising the dynamical equations (\ref{eq:Ph})-(\ref{eq:POP}). On the other hand, the calligraphic letters $\mathcal{A}_3$ and $\mathcal{B}_3$ denote normally hyperbolic equilibrium sets. Note that $\alpha=1/2$ does not appear in table \ref{tab:fpModelKGB3} since the scalar field does not contribute to the evolution of the system. The case of $\alpha=0$ is not present in table \ref{tab:fpModelKGB3} either as it reduces to the model (\ref{model:pureG}) with $\beta=-1/2$ plus a cosmological constant ($\Lambda=-c_K$) and, therefore, the dynamical structure is qualitatively equivalent to that presented in the corresponding column of table \ref{tab:fpBraiding}. 
	
	The events labelled as A$_3$, B$_3$, C$_3$ and $\mathcal{A}_3$ represent vacuum solutions. The latter two belong to group 1, whereas the former two are members of group 2. Moreover, D$_3$ (group 3) admits the same physical interpretation as for D$_1$ and D$_2$; compare with tables \ref{tab:fpkessence} and \ref{tab:fpBraiding}, respectively.
	
	On the other hand, equations (\ref{eq:Jsym}), (\ref{eq:JsymMod3}) and (\ref{eq:rhoMod3}) allow for re-expressing the scalar field energy density as a function of $a$ and $H$ only. Since this is the dominant component at E$_3$ (group 3), and expanding $\rho_\phi$ for large values of the Hubble rate, the Friedmann equation (\ref{FE1}) simplifies as
	\begin{align}\label{eq:HapproxMod3}
		H\approx{\lambda_3}\left(\frac{a}{a_0}\right)^{-\frac{6\alpha}{4\alpha-1}},
	\end{align}
	being ${\lambda_3}\coloneqq (1/3)^\frac{2\alpha-1}{4\alpha-1}[H_0\gamma/(2\alpha-1)c_K]^\frac{1}{4\alpha-1}(\sqrt{2}Q_0/\epsilon)^\frac{2\alpha}{4\alpha-1}$ a positive constant (see section \ref{sec:positivityED}). The exponent above becomes positive when $\alpha\in(0,1/4)$, which signals a future BR singularity. In fact, E$_3$ behaves as an attractor in the phase-space for those values, see table \ref{tab:fpModelKGB3}. Conversely, for $\alpha\in(1/4,1)$ the scalar field dominates over radiation in the past. In that case, E$_3$ represent a scalar-field-induced BB. For the other values of $\alpha$, this equilibrium point features saddle configurations that can be interpreted in the same way as for D$_3$. 
	
	Radiation-induced BB singularity corresponds to F$_3$ and $\mathcal{B}_3$, where for the latter the scalar field contributes to the total radiation content of the universe. These equilibrium points are part of group 4 in our previous classification.
	
	A new feature of this model is the presence of $h$ in the functions $C_1$ and $C_2$. This allows for different solutions corresponding to future dS states (in contrast with the previously discussed models where only one dS attractor was found for a specific value of the corresponding parameter; see tables \ref{tab:fpkessence} and \ref{tab:fpBraiding}). These equilibrium points (group 5) correspond to different solutions to $C_1^{\textup{fp}}=C_2^{\textup{fp}}=0$ and $\Omega_\phi^{\textup{fp}}=1$. According to the structure of the function $C_1$, one such possibility is when the first parenthesis in the numerator of equation (\ref{eq:C1KGBmod3}) cancels. This occurs at
	\begin{eqnarray}\label{eq:h1Mod3}
		h_{1}=\frac{2\alpha-1}{4\alpha\gamma}\left[1\pm\sqrt{1+\left(\frac{4\alpha\gamma}{2\alpha-1}\right)^2}\right].
	\end{eqnarray}
	However, only the negative branch in the preceding expression satisfies the restriction $h_1\in(0,1)$ when $\gamma>0$ and $\alpha\in(0,1/2)$, or $\gamma<0$ and $\alpha\not\in[0,1/2]$. This dS solution has been labelled as $\mathcal{S}_3^1$ in table \ref{tab:fpModelKGB3}.
	
	A second dS solution is that when the second parenthesis in the numerator of $C_1$ vanishes. That leads to
	\begin{align}\label{eq:h2Mod3}
		h_2= \frac{1}{2\gamma}\left(1\pm\sqrt{1+4\gamma^2}\right),
	\end{align}
	where only the negative branch belong to the physical phase-space, i.e. $h_2\in(0,1)$, if $\gamma$ is negative. This event has been assigned the label $\mathcal{S}_3^2$ in table \ref{tab:fpModelKGB3}.
	
	As for the model (\ref{model:pureG}), the dynamical system is potentially ill-defined when the denominator of the auxiliary functions $C_1$ and $C_2$ vanishes. In an expanding universe, it can be shown analytically that this is not the case when $\alpha<0$ or $\alpha>1/2$ if $\gamma$ is positive. However, restricting $\gamma$ to be positive may not always be viable given that $\epsilon$ could change its sign during the evolution of the system when the scalar field is subdominant. (If the scalar field dominates, this parameter can be unambiguously fixed; see section \ref{sec:positivityED}.) Unfortunately, the complexity of the denominator in equations (\ref{eq:C1KGBmod3}) and (\ref{eq:C2KGBmod3}) does not admit an analytic analysis with the background equations like we have done for the model (\ref{model:pureG}). Consequently, fixed points of the system in the region of the phase-space where this denominator cancels, if any, cannot be addressed within this formulation.

	
	\subsection{Conditions for the positivity of the scalar field energy density\label{sec:positivityED}}
	
	In this section we discuss the restrictions on $c_K$, $c_G$ and $\epsilon$ for the energy density (\ref{eq:rhoMod3}) to be positive at least when the scalar field is dominant. In practice, we focus our analysis to the evolution of the system around the equilibrium points where $\Omega_\phi^{\textup{fp}}=1$ in table \ref{tab:fpModelKGB3}. The results are summarized in table \ref{tab:positivityED}.
	
	\paragraph{Vacuum solutions.} These are characterized by $h^{\textup{fp}}=0$. Hence, demanding the scalar field energy density (\ref{eq:rhoMod3}) to be positive when $H\approx0$ implies $(2\alpha-1)c_K>0$. This condition applies to B$_3$ and $\mathcal{A}_3$ in table \ref{tab:fpModelKGB3}. In addition, comparing expressions (\ref{eq:Jsym}) and (\ref{eq:JsymMod3}), it follows that $\epsilon\alpha c_K$ and $Q_0$ have the same sign near these equilibrium points. Therefore, $\epsilon$ is not allowed to change its sign in the nearby configuration space.
	
	\paragraph{Big bang and big rip solutions.} These fixed points correspond to $h^{\textup{fp}}=1$, i.e. $H^{\textup{fp}}\to\infty$. Therefore, the condition $(2\alpha-1)\epsilon c_G>0$ is necessary for the energy density (\ref{eq:rhoMod3}) to be positive when the Hubble rate diverge. This constraint applies to E$_3$ and $\mathcal{B}_3$ in table \ref{tab:fpModelKGB3}. Consequently, $\epsilon$ cannot change its sign in the phase-space around these equilibrium points. Moreover, from comparing equations (\ref{eq:Jsym}) and (\ref{eq:JsymMod3}) it follows that $(2\alpha-1)c_G$ and $Q_0$ should have the same sign. Hence, the constant $\lambda_3$ appearing in equation (\ref{eq:HapproxMod3}) is positive.

	\paragraph{De Sitter solutions.} These are the equilibrium points S$_3^1$ and S$_3^2$ in table \ref{tab:fpModelKGB3}. In order to constrain $c_K$, $c_G$ and $\epsilon$ we impose $J=0$ and $\rho_\phi>0$ in equations (\ref{eq:JsymMod3}) and (\ref{eq:rhoMod3}), respectively. Taking also into account that $H>0$ in an expanding universe, the results are shown in table \ref{tab:positivityED}. Note that we have limited our analysis to those values of the parameter $\alpha$ for which the corresponding dS solutions act as attractors in the configuration space. This is to ensure that the trajectories in the phase-space approach these solutions. 

	\begin{table*}
		\setlength\arrayrulewidth{0.6pt}
		\renewcommand{\arraystretch}{1.5}
		\begin{adjustwidth}{-.65cm}{}
		\begin{tabular}{l c c c c c}
			\hline \hline 
			Fixed Point &  $\alpha<0$ & $0<\alpha<\frac12$ & $\frac12<\alpha<1$ & $\alpha=1$ & $1<\alpha$ \\ \hline
			B$_3$ (vacuum) &$(c_K<0)$ & | & | & | & | \\
			E$_3$ (BB/BR) & | & $(\epsilon c_G<0)$& $(\epsilon c_G>0)$ & | & | \\ \noalign{\vskip 2mm}
			S$_3^1$ (dS) & \parbox{3cm}{$\left\lbrace \left(c_K>0,\ \epsilon c_G<0\right)\right.$, \\  $\left.\left(c_K<0,\ \epsilon c_G>0\right)\right\rbrace$} & $\left(c_K<0,\ \epsilon c_G<0\right)$ & $\left(c_K<0,\ \epsilon c_G>0\right)$ & $\left(c_K<0,\ \epsilon c_G>0\right)$ & $\left(c_K<0,\ \epsilon c_G>0\right)$ \\\noalign{\vskip 2mm}
			S$_3^2$ (dS) & | &  \parbox{3cm}{$\left\lbrace\left(c_K>0,\ \epsilon c_G<0\right)\right.$, \\  $\left.\left(c_K<0,\ \epsilon c_G>0\right)\right\rbrace$}  & | & | & | \\ 
			$\mathcal{B}_3$ (BB) & | & | & | & $(\epsilon c_G>0)$ & |  \\
		\end{tabular}
		\end{adjustwidth}
		\caption{Necessary constraints for the scalar field energy density (\ref{eq:rhoMod3}) to be positive near the equilibrium points dominated by the scalar field. The conditions on $c_K$, $c_G$ and $\epsilon$ are shown only when the corresponding fixed point acts as an attractor or a repeller in the configuration space, i.e. when trajectories in the phase-space undoubtedly approach or move away from that solution. Note that E$_3$ is not a fixed point when $\alpha=1/4$ (see table \ref{tab:fpModelKGB3}). A set of conditions is showed in parenthesis. Multiple possible sets of conditions are grouped in curly brackets. Parameters that are not mentioned remain unrestricted.\label{tab:positivityED}}
	\end{table*}

	
	\section{Conclusions\label{sec:conclusions}} 
	
	The possibility of a stable self-tuning dS attractor in the shift-symmetric KGB theories has naturally attracted the attention of the scientific community \cite{DeFelice:2010pv,Bernardo:2021hrz,Germani:2017pwt,Martin-Moruno:2015bda,Martin-Moruno:2015kaa,DeFelice:2011bh,Tsujikawa:2010zza}. Furthermore, revising the literature it could seem that this is
	the only possible future evolution for these cosmological models. Nevertheless,	different future evolutions are also possible. In order to analyse this issue, we have proposed a dynamical system formulation never applied before to shift-symmetric KGB theories. The key feature of this new formulation is the compactification of the Hubble rate in the configuration space. This allows for the proper identification of cosmological singularities where the Hubble rate and its cosmic time derivative diverge but the ratio $\dot{H}/H^2$ is finite (i.e $C_1$ finite) as fixed points of the system. The physical interpretation of these cosmological singularities may vary depending on when and where they take place in the phase-space.
	
	Owing to the structure of the dynamical equations (\ref{eq:Ph})-(\ref{eq:POP}), we have found at least five different groups of fixed points. (Recall that there may be equilibrium points that have eluded this classification because of the choice of the dynamical variables being not adequate for them to be correctly identified.) The existence and stability of these fixed points, however, depends ultimately on the choice of the functions $K$ and $G$ of the system. In sections \ref{sec:powerLaws} and \ref{sec:Mod3}, we have applied this description to different power law examples. For these power law functions different future cosmic singularities acting as attractors in the corresponding configuration space have been identified. Most notoriously, having a future evolution towards a BR singularity was found to be always possible for the proposed models. This is (to the best of our knowledge) the first time this cosmic singularity has been explicitly found in the shift-symmetric KGB sector. (See also the companion reference \cite{TMPletter}.) Our findings advocate for a richer future phenomenology of the KGB theories than previously expected. Indeed, we consider this broader future phenomenology to significantly contribute to the interest of shift-symmetric KGB models in cosmology.
	
	Additionally, we have also identified the occurrence of BF and sudden singularities for the KGB model (\ref{model:pureG}). The presence of finite-size singularities provides an excellent example why $J=0$ is not, in general, an exhaustive characterization of all the possible future attractors for an expanding universe in the shift-symmetric KGB theory. This is because the observable universe reaches a maximum size and, therefore, the shift-current  (\ref{eq:Jsym}) is non-trivial on the attractor as long as the shift-charge $Q_0$ is not null.
	
	It should be highlighted, however, that the analysis of the background cosmic evolution we have performed here must be combined with a discussion on the stability of the cosmological perturbations in order to address the viability of the KGB models under consideration.
	The conditions for the absence of ghost and gradient instabilities for scalar perturbations were already obtained in references  \cite{KGB1,DeFelice:2011bh} (see also references \cite{Helpin:2019kcq,Bellini:2014fua}). Therefore, the fulfilment of these conditions at least at the vicinity of the fixed points obtained in tables \ref{tab:fpkessence} to \ref{tab:fpModelKGB3} should be considered as a necessary but not sufficient condition for the stability of the scalar perturbations during the whole evolution. At the vicinity of the scalar field dominated fixed points with a phantom equation of state ($w_\phi^{\textup{fp}}<-1$) we have found here, which are the main results of our approach, the ghost and/or gradient condition are always violated. This may signal that the phantom solutions we have discussed are not viable from the point of view of the scalar perturbations. 
	Nevertheless, it would be worthwhile to investigate whether the braiding term could lead to a (non-trivial) non-adiabatic regime in perturbation theory. This could be feasible due to the presence of the Hubble rate in the energy density of the scalar field. Therefore, if that is the case, it would be interesting to explore whether the non-adiabatic perturbations could contribute alleviating the instabilities of these phantom models. A similar discussion for a phantom DE model with a future BR singularity can be found in reference \cite{Albarran:2016mdu}.

	
	\section*{Acknowledgements}
	
	The research of T.B.V. and P.M.M. is supported by MINECO (Spain) Project No. PID2019-107394GB-I00 (AEI/FEDER, UE). 
	T.B.V. also acknowledge financial support from Universidad Complutense de Madrid and Banco de Santander through Grant No. CT63/19-CT64/19. He is also grateful for the hospitality of the University of the Basque country (UPV-EHU) where this work was partly developed.
	The work of M.B.L. is supported by the Basque Foundation of Science Ikerbasque. Her work has been also financed by the Spanish project PID2020-114035GB-100 (MINECO/AEI/FEDER, UE). She would like to acknowledge the financial support from the Basque government Grant No. IT1628-22 (Spain). 
	
	
	\appendix
	
	\section{Auxiliary functions\label{app:functions}}
	
	Taking into account the expression for the shift-current (\ref{eq:J}), the scalar field equation (\ref{eq:phi}) can be expanded as
	\begin{align}
		&A(H,X)X'+6\epsilon\sqrt{2}X^\frac32 G_X H'+3\epsilon\sqrt{2X}J=0,\label{phi.bis}
	\end{align}	
	with the function
	\begin{align}
		A(H,X)\coloneqq& K_X+2XK_{XX}+6\epsilon\sqrt{2X}H\left(G_X+XG_{XX}\right),
	\end{align}
	introduced for the sake of the notation. In addition, the Raychaudhuri equation (\ref{FE2}) can be re-expressed as
	\begin{align}
		&HH'-3\epsilon\sqrt{2X}G_XHX'+9H^2+\rho_r+3K=0,\label{FE2.bis}
	\end{align}
	where the Friedmann equation (\ref{FE1}) has been used to eliminate $\rho_m$.
	Thus, expressions (\ref{phi.bis}) and (\ref{FE2.bis}) can be thought of as a system of two equations for $H'$ and $X'$. Provided that this system of equations is non-degenerated, the solutions are
	\begin{align}
		H'=&-\frac{(9H^2+\rho_r+3K) A+18XG_XHJ}{6H\left(A+6X^2G_X^2\right)},\\
		X'=& \frac{\epsilon\sqrt{2X}\left[\left(9H^2+\rho_r+3K\right)XG_X-3HJ\right]}{H\left(A+6X^2G_X^2\right)}.
	\end{align}
	Hence, the auxiliary functions $C_1$ and $C_2$ read
	\begin{align}
		C_1=&-\frac{(9H^2+\rho_r+3K) A+18XG_XHJ}{6H^2\left(A+6X^2G_X^2\right)},\\
		C_2=&\frac{2XG_X\left[XG_X(9H^2+\rho_r+3K)-3HJ\right] }{H^2\left(A+6X^2G_X^2\right)}-\frac{\epsilon\sqrt{2X}J}{H^2},
	\end{align}
	see definitions in (\ref{def:C1}) and (\ref{def:C2}), respectively. Note that these functions depend on $H$, $X$ and $\rho_r$ but not on their time derivatives.	Once $K(X)$ and $G(X)$ are specified, these functions can be completely re-written in terms of the new variables $h$, $\Omega_r$ and $\Omega_\phi$ if definition (\ref{varOP}) can be inverted to obtain the kinetic term $X$ as a function on $h$ and $\Omega_\phi$.
	
	
	\section{Big rip singularity\label{app:BR}}
	
	This appendix is to justify and remind that a future BR singularity \cite{BR,Starobinsky:1999yw} takes place when the Hubble rate is proportional to a positive power of the scale factor. Lets assume that for $a\geq a_\star$, with $a_\star$ some reference scale, we have
	\begin{eqnarray}\label{eq:exampleBR}
		H(a)\approx \lambda \left(\frac{a}{a_0}\right)^p,
	\end{eqnarray}
	being $\lambda$ and $p$ positive constant. The scale factor, then, evolves in time as
	\begin{eqnarray}
		a(t)=a_0\left[\frac{1}{p\lambda(t_r-t)}\right]^\frac1p,
	\end{eqnarray}
	where 
	\begin{eqnarray}
		t_r\coloneqq t_\star+ \frac{1}{p\lambda}\left(\frac{a_0}{a_\star}\right)^p.
	\end{eqnarray}
	Note that $t_r>t_\star$ since $\lambda$ and $p$ are positive. Hence, the scale factor diverge at some finite future moment $t_r$. Similarly, the Hubble rate and its cosmic time derivative also blow-up at $t_r$ given that
	\begin{align}
		H(t)&=\frac{1}{p(t_r-t)},\\
		\dot{H}(t)&=\frac{1}{p(t_r-t)^2}.		
	\end{align}
	Therefore, a future BR singularity takes place at $t=t_r$.
	

\end{document}